\documentclass[twocolumn]{aastex631}
\usepackage{amsmath}
\usepackage{amssymb}
\usepackage{apjfonts}

\newlength{\apjcolwidth}
\begin{document}

\title{Phases of Mass Transfer from Hot Subdwarfs to White Dwarf
  Companions and Their Photometric Properties}

%% Note that the corresponding author command and emails has to come
%% before everything else. Also place all the emails in the \email
%% command instead of using multiple \email calls.
\correspondingauthor{Evan B. Bauer}
\email{evan.bauer@cfa.harvard.edu}

\author[0000-0002-4791-6724]{Evan B. Bauer}
\affiliation{Center for Astrophysics $\vert$ Harvard \& Smithsonian, 60 Garden St, Cambridge, MA 02138, USA}
\affiliation{Kavli Institute for Theoretical Physics, University of California, Santa Barbara, CA 93106, USA}

\author[0000-0002-6540-1484]{Thomas Kupfer}
\affiliation{Department of Physics and Astronomy, Texas Tech University, PO Box 41051, Lubbock, TX 79409, USA}
\affiliation{Kavli Institute for Theoretical Physics, University of California, Santa Barbara, CA 93106, USA}

\begin{abstract}
  Binary systems of a hot subdwarf B (sdB) star
  + a white dwarf (WD) with orbital periods less than
  2--3 hours can come into contact due to gravitational waves and transfer
  mass from the sdB star to the WD before the sdB star ceases nuclear
  burning and contracts to become a WD. Motivated by the growing class
  of observed systems in this category, we study the phases of mass
  transfer in these systems. We find that because the
  residual outer hydrogen envelope accounts for a large fraction of an
  sdB star's radius, sdB stars can spend a significant amount of time
  ($\sim$10s of Myr) transferring this small amount of material at low
  rates ($\sim 10^{-10}$--$10^{-9}\ M_\odot\,\rm yr^{-1}$)
  before transitioning to a phase where the bulk of
  their He transfers at much faster rates ($\gtrsim 10^{-8}\ M_\odot\,\rm yr^{-1}$).
  These systems therefore spend a surprising amount of time with Roche-filling sdB
  donors at orbital periods longer than the range associated with He
  star models without an envelope.
  We predict that the envelope transfer phase should be
  detectable by searching for ellipsoidal modulation of Roche-filling
  objects with $P_{\rm orb}=30$--$100$~min and
  $T_{\rm eff}=20{,}000$--$30{,}000$~K, and that many ($\geq$10) such
  systems may be found in the Galactic plane after accounting for
  reddening. We also argue that many of these systems may go through a
  phase of He transfer that matches the signatures of
  AM CVn systems, and that some AM CVn systems associated with
  young stellar populations likely descend from this channel.
\end{abstract}


\section{Introduction}

Hot subdwarf B (sdB) stars are subluminous stars of spectral type B,
thought to be compact He-burning stars of mass
$\approx$0.3--0.6~$M_\odot$ with thin hydrogen envelopes
\citep{Heber2009,Heber2016,Zhang2009,Gotberg2018,Geier2020}. The low masses of these He-burning
objects that have not yet gone through AGB mass loss suggest
that interaction with a binary companion is needed to explain
the existence of sdB stars \citep{Maxted2001,Han2002,Pelisoli2020}.
Indeed, in many cases sdB stars are observed in compact binary systems
with orbital periods of only a few hours, implying a prior common
envelope event that ejected most of the sdB star's former H envelope
and caused the companion star to spiral inward toward its current
orbital configuration
\citep{Han2002,Han2003,Napiwotzki2004,Nelemans2010,Copperwheat2011,Kruckow2021}. The most
compact of these systems consist of sdB
(or sometimes even hotter sdO)
stars with white dwarf (WD)
companions with orbital periods on the order of just one hour
\citep{Vennes2012,Geier2013,Kupfer2017a,Kupfer2017b,Kupfer2020a,Kupfer2020b,Pelisoli2021},
which can make contact and transfer mass from the sdB/sdO star to
the WD via stable Roche lobe overflow within the sdB/sdO star's He-burning
lifetime \citep{Justham2009,Brooks2015}.\footnote{For convenience in
  the remainder of this paper, we will use the shorthand ``sdB'' to
  refer to the hot subdwarfs of mass $\approx$0.3--0.6~$M_\odot$ that
  we model in this work, though in some stages models may also evolve
  toward hotter temperatures where observations may classify them as
  sdOB or sdO stars.}

Evolution of these compact sdB+WD binaries can lead to many
interesting astrophysical phenomena, including thermonuclear events
due to accretion on the WD and runaway subdwarf remnants in the case of a
supernova that disrupts the binary
\citep{Hirsch2005,Justham2009,Wang2009,Piersanti2014,Geier2015,Brooks2015,Neunteufel2016,Neunteufel2019,Bauer2017,Bauer2019,Neunteufel2020}.
Some observed thermonuclear transients show evidence of thick He shell
detonations \citep{Polin2019,De2019,De2020}, consistent with the
binary evolution predictions for mass transfer from sdB stars to WD
companions.

These binaries may also be important gravitational wave sources for the
Laser Interferometer Space Antenna (LISA), motivating work toward
better characterization of the population of these binaries in our
Galaxy \citep{Gotberg2020}. For systems in which the sdB star is at or
near Roche filling, high-cadence photometry can detect gravity darkening and ellipsoidal
modulation of the lightcurve due to temperature variations on the
distorted subdwarf surface, as \cite{Kupfer2020a,Kupfer2020b}
recently demonstrated in discovering two new Roche-filling systems
with subdwarf donor stars.

Although the hydrogen envelope of an sdB star contains only a small
fraction of the star's total mass
($M_{\rm env} \sim 10^{-4} - 10^{-2}\ M_\odot$), the envelope extends over a
significant fraction of the total radius, and therefore has a
large impact on the range of orbital periods for which sdB stars
are Roche-filling objects that display significant ellipsoidal
modulation. Previous work has emphasized the $\approx$20--40 minute
orbital period range for compact Roche-filling sdB cores calculated
neglecting the H envelope
\citep{Savonije1986,Iben1991,Piersanti2014,Brooks2015}. In this paper,
we use the stellar evolution code  Modules for
Experiments in Stellar Astrophysics (MESA) to calculate models that
explore the sensitivity of sdB radii
to H envelope masses and map out the space of possible orbital periods
where Roche-filling mass transfer can occur. We find that when
accounting for the H envelope, sdB stars can fill their Roche lobes at
longer orbital periods and for a much longer portion of their lifetimes.
This has significant
implications for the potential range of orbital periods at which these
systems can be discovered by searching for ellipsoidal modulation in
high-cadence photometry.

In Section~\ref{s.models}, we calculate a grid of MESA sdB star models
to demonstrate the dependence of the radius on the thin H envelope
mass.
In Section~\ref{s.analytics}, we describe the physics of mass transfer
for Roche-filling sdB stars, and we show that the mass-radius
relations from our MESA models are sufficient to predict accretion
rates and timescales through both the H and He mass transfer
phases. Surprisingly, the H mass transfer phase can last for $>10$~Myr,
as long or longer than the subsequent He mass transfer phase.
In Section~\ref{s.periods}, we use a grid of MESA sdB models to map
out the space of orbital periods for which systems can fill their
Roche lobes and begin transferring mass. We find that H envelope mass
transfer can begin at orbital periods as long as $\approx$100 minutes,
and that 40--60~minutes may be a typical period for the onset of mass transfer.
In Section~\ref{s.MESAbinaries}, we verify our calculations with two
full binary evolution models for sdB+WD systems in which we model both
stars along with the orbital evolution and mass transfer using MESA.
  In Section~\ref{s.ellipsoidal}, we demonstrate that the ellipsoidal
  variation for these sdB+WD systems will be easily detectable for
  many systems within a few kpc, which will enable discovery of many
  new systems.
Finally, in Section~\ref{s.conclusions} we discuss the implications of
our work for targeting searches for these binary systems based on
ellipsoidal modulation, and predict that at least 10s of these systems may be
discovered within a few kpc in the Galactic plane after accounting for reddening.

%\newpage
\section{sdB Star Radius as a Function of Envelope Mass}
\label{s.models}

As we will show in Section~\ref{s.analytics}, the first phase of mass
transfer from sdB star donors onto WD companions depends primarily on
the sdB star envelope mass-radius relation, so in this section we use
grids of MESA models to show this relation for use in later sections.
Our MESA models all use release version 15140
\citep{Paxton2011,Paxton2013,Paxton2015,Paxton2018,Paxton2019}.
The MESA equation of state (EOS) is a blend of the OPAL \citep{Rogers2002}, SCVH
\citep{Saumon1995}, FreeEOS \citep{Irwin2004}, HELM \citep{Timmes2000},
and PC \citep{Potekhin2010}.
Radiative opacities are primarily from OPAL \citep{Iglesias1993,
Iglesias1996}, and electron conduction opacities are from
\citet{Cassisi2007}.
Nuclear reaction rates are from JINA REACLIB \citep{Cyburt2010} plus
additional tabulated weak reaction rates \citet{Fuller1985, Oda1994,
Langanke2000}.
Screening is included via the prescription of \citet{Chugunov2007}.
Thermal neutrino loss rates are from \citet{Itoh1996}.

Our MESA sdB models employ the ``predictive'' mixing scheme
\citep{Paxton2018} for convection in the He-burning core. This allows
the convective core to grow over the sdB star lifetime
without requiring convective overshoot \citep{Schindler2015} while
also providing a setting that prevents the core from growing too far
and dividing, which can cause ``breathing pulses'' that cause the core
size to fluctuate along with the star luminosity and radius. Breathing
pulses would greatly complicate the study of mass transfer via
Roche-lobe overflow from sdB stars, but they are likely numerical
artifacts rather than physical phenomena \citep{Paxton2019}, so
we use the predictive mixing scheme so that our models avoid them
during core He burning. For a more thorough discussion of convection
treatments in MESA sdB models, see \cite{Ostrowski2020}.

\begin{figure}
  \centering
  \includegraphics{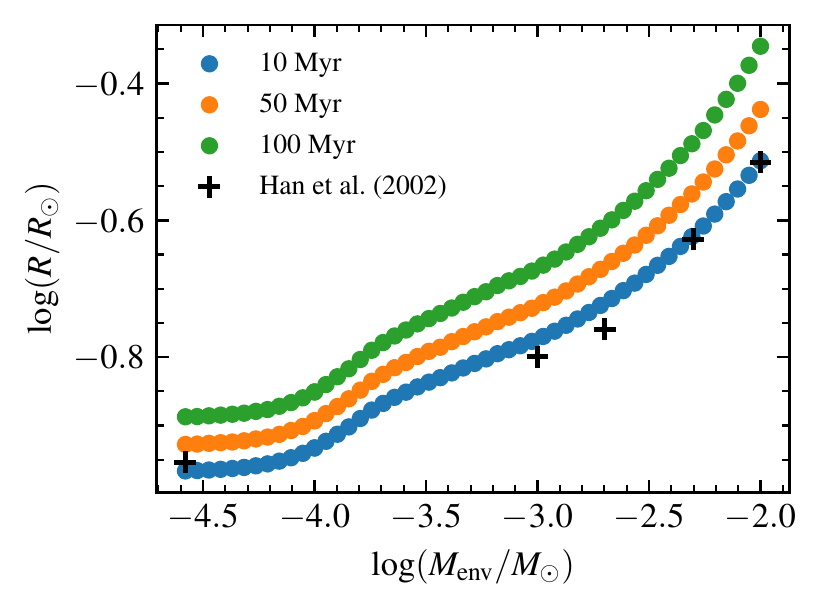}
  \caption{Grid of MESA models showing sdB star radius as a function
    of hydrogen envelope mass for a 0.47~$M_\odot$ sdB star. Each
    point of a given color represents a distinct MESA model in the grid, with colors
    indicating sdB star age. For comparison, the black crosses show
    estimated sdB radii based on the beginning of the $\log
    g$--$T_{\rm eff}$ tracks of \cite{Han2002} for 0.46~$M_\odot$ sdB
    models.}
  \label{fig:sdbMR}
\end{figure}

Our first grid of MESA sdB models represents a canonical
0.47~$M_\odot$ sdB star with a range of envelope masses.
These models
are constructed from a $1\ M_\odot$ progenitor star with metallicity
$Z=0.02$ that undergoes the He core flash when the core reaches a mass
of $0.47\ M_\odot$. After the He core flash, we remove all but a thin
residual H envelope, and evolve the models as sdB stars to produce
the relation between sdB radius $R$ and envelope mass $M_{\rm env}$
shown in Figure~\ref{fig:sdbMR}. Envelope mass $M_{\rm env}$ is
defined here as all the mass of material contained in the exterior
layers for which the H mass fraction is $X > 0.01$. For this set of
0.47~$M_\odot$ models, the envelope material has roughly solar
composition matching the initial ZAMS composition of the progenitor
$1\ M_\odot$ star.
  Figure~\ref{fig:sdbMR} shows three snapshots at different ages
  along the radius evolution in this grid of models, showing that
  these sdB models slowly evolve from smaller to larger radii for most
  of their core He burning lifetime.
Figure~\ref{fig:sdbMR} also shows radius
estimates from the models of \cite{Han2002} based on the beginning of
the $\log g$--$T_{\rm eff}$ tracks in figure~2 of that
paper, confirming that our models generally agree with another stellar
evolution code commonly used to model sdB stars.
Envelope masses $\log(M_{\rm env}/M_\odot) < -4.5$ in
Figure~\ref{fig:sdbMR} give radii that converge toward the radius of a
model with a bare He core and no envelope.

\begin{figure}
  \centering
  \includegraphics{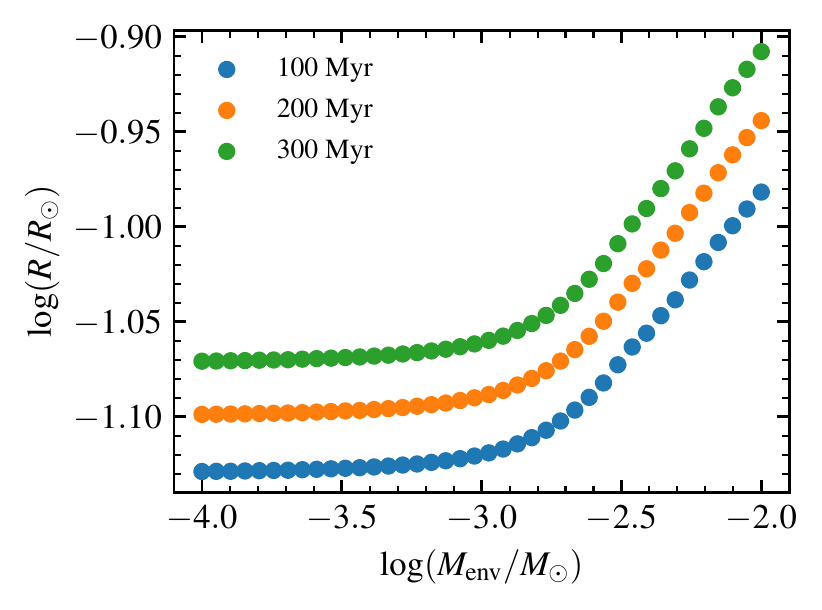}
  \caption{Same as Figure~\ref{fig:sdbMR} but for a 0.34~$M_\odot$ sdB
    star.}
  \label{fig:sdbMR_lighter}
\end{figure}

To illustrate the case of a less massive sdB star donor, we also run a
second grid of MESA models for a 0.34~$M_\odot$ sdB star.
This smaller He-burning
core is descended from a $2.8\ M_\odot$ progenitor star that does not
go through the He core flash, and the residual H-rich material that
forms the sdB envelope is enriched in He due to partial nuclear
processing of that region in the receding convective core on the main
sequence (see \citealt{Kupfer2020a} for more details).
Figure~\ref{fig:sdbMR_lighter} shows the $M_{\rm env}$-$R$ relation
for this lower-mass grid of MESA sdB models. Envelope masses
$\log(M_{\rm env}/M_\odot) < -3$ for this model yield radii that
quickly converge to the radius of a bare He core. This is due to the
smoother composition transition between the He core and H envelope,
which leaves only a few percent H by mass in the residual outer
envelope when $\log(M_{\rm env}/M_\odot) < -3$.

\section{Mass Transfer Physics}
\label{s.analytics}

We now present the physics that describes the
mass transfer rate for both the envelope and He-rich core of the sdB star.
In Section~\ref{s.MESAbinaries}, we will also compare these mass transfer
rate predictions to some example MESA binary evolution models, and
confirm that they show general agreement with the analytic calculations
here.

We label the sdB as star 1 with radius $R_1$ and mass $M_1$,
and the WD as star 2 with radius $R_2$ and mass $M_2$.
The mass ratio is $q \equiv M_1/M_2$, and following
\cite{Eggleton1983}, we approximate the Roche radius for the hot
subdwarf as
\begin{equation}
  R_{RL} = \frac{0.49q^{2/3}a}{0.6q^{2/3} + \ln(1 + q^{1/3})}~,
  \label{eq:EggletonRL}
\end{equation}
where $a$ is the orbital separation between the two stars.
Kepler's third law gives the relation
between orbital separation and period:
\begin{equation}
  a^3 = \frac{G(M_1 + M_2)P_{\rm orb}^2}{4 \pi^2}~.
  \label{eq:Kepler3}
\end{equation}

When the sdB star begins to overflow its Roche lobe, the mass transfer
rate depends on the response of the star's radius to losing mass,
which can be described in terms of an exponent $\zeta$ for the
relation $R_1 \propto M_1^\zeta$ \citep{Soberman1997}.
If mass transfer is slow enough that
the star can maintain thermal equilibrium, the relevant exponent is
\begin{equation}
  \zeta_{\rm eq} \equiv \left(\frac{d \ln R_1}{d \ln M_1}\right)_{\rm eq}~,
\end{equation}
which simply encodes the mass-radius relation for a star in
equilibrium with no structural effects from losing mass.
On the other hand, if thermal transport is not efficient
enough to maintain equilibrium on the timescale of mass removal, the
entropy profile in the stellar interior will remain fixed, so that the
relevant exponent encodes the adiabatic radius response:
\begin{equation}
  \zeta_{\rm ad} \equiv \left(\frac{d \ln R_1}{d \ln M_1}\right)_{\rm ad}~.
\end{equation}

The overall timescale for orbital evolution is set by the rate at which
gravitational-wave radiation (GWR) shrinks the orbital separation and
Roche lobe:
\begin{equation}
  \tau_{\rm gr} = \frac{J_{\rm orb}}{|{\dot J}_{\rm gr}|}~,
\end{equation}
where
\begin{equation}
  \begin{split}
    {\dot J}_{\rm gr}
    & = - \frac{32}{5c^5}\left(\frac{2 \pi G}{P_{\rm orb}}\right)^{7/3}
    \frac{(M_1 M_2)^2}{(M_1 + M_2)^{2/3}} \\
    &= - \frac{32}{5c^5} \left( \frac{G}{a} \right)^{7/2}
    (M_1 M_2)^2 \sqrt{M_1 + M_2}~,
  \end{split}
  \label{eq:Jdotgr}
\end{equation}
and
\begin{equation}
  J_{\rm orb} = M_1 M_2 \sqrt{\frac{Ga}{M_1 + M_2}} ~.
  \label{eq:Jorb}
\end{equation}

In order to estimate the mass transfer rate, it is also necessary to
account for the change in the Roche radius that results 
from transferring mass from star 1 to star 2:
\begin{equation}
  \zeta_{RL} \equiv \frac{d \ln R_{RL}}{d \ln M_1}
  = \frac{d \ln a}{d \ln M_1} +
  \frac{d \ln(R_{RL}/a)}{d \ln q} \frac{d\ln q}{d\ln M_1}~.
\end{equation}
For conservative mass transfer ($M_1+M_2 = \text{constant}$) and
constant $J_{\rm orb}$, we can use Equation~\eqref{eq:Jorb} to
evaluate the first term on the right hand side as
$d\ln a/d\ln M_1 = 2(q-1)$. Conservative mass transfer also yields
$d\ln q/d\ln M_1 = 1+q$. Differentiating Equation~\eqref{eq:EggletonRL} gives
\begin{equation}
  \frac{d \ln(R_{RL}/a)}{d \ln q} =
  \frac{q^{1/3}}{3}\left(\frac{2}{q^{1/3}}
    - \frac{1.2q^{1/3} + \frac{1}{1+q^{1/3}}}{0.6q^{2/3} + \ln(1+ q^{1/3})} \right)~.
\end{equation}
% For an estimate of the second term, it
% is convenient to use the \cite{Paczynski1971} approximation for the
% Roche lobe radius:
% \begin{equation}
%   \frac{R_{RL}}{a} \approx 0.46224 \left( \frac{M_1}{M_1+M_2} \right)^{1/3} ~,
% \end{equation}
% which is accurate to within 2\% for $q<0.8$ (applicable for almost any
% system considered in this paper), and gives $d \ln(R_{RL}/a)/d \ln M_1
% \approx 1/3$ for conservative mass transfer.
So the full expression for the Roche radius response to conservative
mass transfer can be written as
\begin{equation}
  \zeta_{RL}
  = 2(q-1) +
    \frac{(1+q)q^{1/3}}{3}\left(\frac{2}{q^{1/3}}
    - \frac{1.2q^{1/3} + \frac{1}{1+q^{1/3}}}{0.6q^{2/3} + \ln(1+ q^{1/3})} \right)\,.
  \label{eq:zetaRLfull}
\end{equation}
It will often be convenient to adopt the much simpler approximation
\begin{equation}
  \zeta_{RL} \approx - \frac 5 3 + 2.1q~,
  \label{eq:zetaRLest}
\end{equation}
which can be shown to agree with Equation~\eqref{eq:zetaRLfull} to
within a few percent.
Note that $\zeta_{RL} < 0$ for $q \lesssim 0.8$, so in general the Roche lobe
radius expands in response to mass lost from the sdB donor for the systems
with $M_{\rm WD} \geq 0.6\ M_\odot$ C/O WD accretors that are the
focus of this paper.

Lastly, we also must account for the timescale for the donor star's
radius to change due to stellar evolution independent of mass transfer
\begin{equation}
  \tau_R = \left( \frac{d \ln R_1}{dt} \right)^{-1} ~.
  \label{eq:tauR}
\end{equation}
We assume fully conservative mass transfer here, so that
$\dot J_{\rm gr}$ is the only term affecting the net angular momentum
of the system, and
the equilibrium mass transfer rate can then be
estimated as (see e.g., \citealt{Ritter1988})
\begin{equation}
  \dot M_1 = \frac{M_1}{\zeta - \zeta_{RL}}
  \left(\frac{2}{\tau_{\rm gr}} + \frac{1}{\tau_R}\right) ~,
  \label{eq:mdot_full}
\end{equation}
where $\zeta$ is determined by comparing the timescale for orbital
evolution to the thermal (Kelvin-Helmholtz) timescale:
\begin{equation}
  \tau_{\rm KH} = \frac{GM_1^2}{R_1L_1}~.
\end{equation}
We use $\zeta = \zeta_{\rm eq}$ when $\tau_{\rm KH} < \tau_{\rm gr}$
and $\zeta = \zeta_{\rm ad}$ when $\tau_{\rm KH} > \tau_{\rm gr}$, and
mass transfer is stable as long as $\zeta_{RL} < \zeta$.

Throughout this paper, we will assume that both stars in the binary
maintain co-rotation with the orbit, so that we can neglect angular
momentum that might be lost from the orbit e.g.\ due to accretion
causing the WD to spin up. Direct impact accretion is thought to
violate this assumption in double white dwarf systems
\citep{Nelemans2001,Marsh2004}, causing dynamically unstable mass
transfer in systems with $q < 2/3$ that would otherwise be
stable. However, the sdB+WD systems that we consider here have larger
orbital separations due to the larger sdB radius when it overflows its
Roche lobe. Using the \cite{Nelemans2001} fit to the \cite{Lubow1975}
calculations for minimum distance of the accretion stream from the
accretor, we find that sdB+WD binary systems will never
come close to the direct impact regime due to the larger orbital
separations required for sdB donors.
These systems are therefore expected to form an accretion
disk that transfers most of its angular momentum back into the orbit
\citep{Verbunt1988,Priedhorsky1988}. Therefore, we will find that mass
transfer is stable in almost all of the systems that we consider here,
except the case of low mass sdB donors ($M_{\rm sdB} \lesssim 0.35\ M_\odot$)
transferring He in systems with $q > 2/3$ (see Section~\ref{s.he_transfer}).

\subsection{Envelope Transfer}

To represent a typical subdwarf, we consider an sdB star of total
mass $M_1 = 0.47\ M_\odot$ with an initial hydrogen envelope mass of
$10^{-3}\ M_\odot$. Figure~\ref{fig:sdbMR} shows that such a subdwarf
will have a radius in the range $R_1 = 0.16$--$0.20\ R_\odot$ for most
of its lifetime. For the example binary evolution scenarios
presented throughout this paper, we will adopt a WD companion mass of $M_2 =
0.75\ M_\odot$, which is representative of many of the short-period
sdB+WD systems that have been observed so far
(e.g., \citealt{Geier2013,Kupfer2020b}; Kupfer et al.~2021, in prep),
likely reflecting the fact that these systems are found among young
populations. For this companion mass $M_2 = 0.75\ M_\odot$,
Equations~\eqref{eq:EggletonRL} and~\eqref{eq:Kepler3} together imply
that the sdB star will initially fill its Roche lobe ($R_1 = R_{RL}$)
at an orbital period of 50--70~min.
For the sdB stars shown in Figure~\ref{fig:sdbMR} with luminosities
on the order of $L \approx 20\ L_\odot$, the Kelvin-Helmholtz
timescale is $\tau_{\rm KH} \lesssim 2\ {\rm Myr}$.
For $M_1 = 0.47\ M_\odot$ and $M_2 = 0.75\ M_\odot$ and $P_{\rm orb} =
50\ {\rm min}$, $\tau_{\rm gr} \approx 150\ {\rm Myr} >> \tau_{\rm KH}$,
and the star's radius response is therefore described by $\zeta_{\rm eq}$.
From the timescale for radius evolution shown in
Figure~\ref{fig:sdbMR}, we can estimate
$\tau_R \approx 400\ \rm Myr$ based on
the change in $\ln R_1$ according to
Equation~\eqref{eq:tauR}.

For mass transfer of the envelope, we can write the star's radius
response in terms of an exponent for the envelope mass rather than the
total stellar mass:
\begin{equation}
  \zeta_{\rm eq,env} = \left(\frac{d \ln R_1}{d \ln M_{\rm env}}\right)_{\rm eq}
  = \frac{M_{\rm env}}{M_1} \zeta_{\rm eq} ~,
\end{equation}
which can be estimated from Figure~\ref{fig:sdbMR}, giving values of
$\zeta_{\rm eq,env} \approx 0.2$--$0.3$.
This implies that $\zeta_{\rm eq} \gg 1$ because $M_1 \gg M_{\rm env}$.
On the other hand, $\zeta_{RL}$ is of order unity according to Equation~\eqref{eq:zetaRLest}
(cf.\ figure 4 in \citealt{Soberman1997}), so $\zeta_{RL}$ can be
neglected when applying Equation~\eqref{eq:mdot_full} to transferring
the envelope. We can therefore approximate the instantaneous mass
transfer rate for the hydrogen envelope as
\begin{equation}
  \dot M_{\rm env}
  \approx \frac{M_{\rm env}}{\zeta_{\rm eq,env}}
  \left(\frac{2}{\tau_{\rm gr}} + \frac{1}{\tau_R} \right)~,
  \label{eq:MdotEnv}
\end{equation}
which predicts that when the subdwarf
first fills its Roche lobe at $P_{\rm orb} = 50$--$70\ \rm min$, mass
transfer will occur at rates in the range of
$\dot M_{\rm env} \approx 2$--$8 \times 10^{-11}\ M_\odot\,{\rm yr}^{-1}$.
This mass transfer is driven primarily by the shrinking of the orbit
due to GWR ($\tau_{\rm gr}$ term), with a smaller contribution from the slowly expanding
radius of the star pushing material past its Roche lobe ($\tau_R$ term).

Without a hydrogen envelope, Figure~\ref{fig:sdbMR} shows that the
bare helium core has a substantially smaller equilibrium radius
on the order of $0.11$--$0.13\ R_\odot$, which fills the Roche lobe at
an orbital period of 28--36~min (see also Figure~\ref{fig:HeMR}).
We can write the GWR merger time of two point masses as
\begin{equation}
  \tau_{\rm merge} = \frac{5}{256} \frac{c^5 a^4}{G^3 M_1 M_2 (M_1 + M_2) }~.
  \label{eq:tmerge}
\end{equation}
This equation describes the time evolution of the system's
orbital separation before the stars are in contact ($M_1$ and $M_2$ constant), and it can be
shown that it also applies as long as the stars are transferring mass
at a rate that is negligible for affecting the angular momentum of
each star compared to GWR evolution. In this context, a rate of
$\dot M \lesssim 10^{-9}\ M_\odot\,\rm yr^{-1}$ is negligible, so
Equation~\eqref{eq:tmerge} applies during the envelope transfer phase
as well as prior to the stars coming into contact. Using this equation
along with Equation~\eqref{eq:Kepler3}, we find that the inspiral time to reach a
period of 30~min from a period of 50~min (70~min) is 14~Myr
(40~Myr). Therefore, after the hydrogen envelope begins to overflow
the Roche lobe, the mass transfer must occur at an average rate on the
order of $2$--$7 \times 10^{-11}\ M_\odot\,{\rm yr}^{-1}$ until the
$10^{-3}\ M_\odot$ envelope is exhausted, which agrees with the
prediction of Equation~\eqref{eq:MdotEnv}.

A less massive sdB star
(like the 0.34~$M_\odot$ models shown in Figure~\ref{fig:sdbMR_lighter})
will not fill its
Roche lobe until reaching an orbital period of $P_{\rm orb} =
25$--$40$~min (for a $0.75\ M_\odot$ WD companion), and the period at
which the bare He core is exposed is $P_{\rm orb} = 18\ \rm min$. The
inspiral timescale according to Equation~\eqref{eq:tmerge} is
therefore only 2--10~Myr, and mass transfer of the envelope must occur
at a rate on the order of
$\dot M_{\rm env} \sim 10^{-10}$--$10^{-9}\ M_\odot\,\rm yr^{-1}$.

Once the hydrogen envelope is exhausted after a few to 10s of Myr, the He-rich
layers underneath will be revealed and mass transfer will proceed at a
much higher rate.

\subsection{The Helium Core}
\label{s.he_transfer}

\begin{figure}
  \centering
  \includegraphics{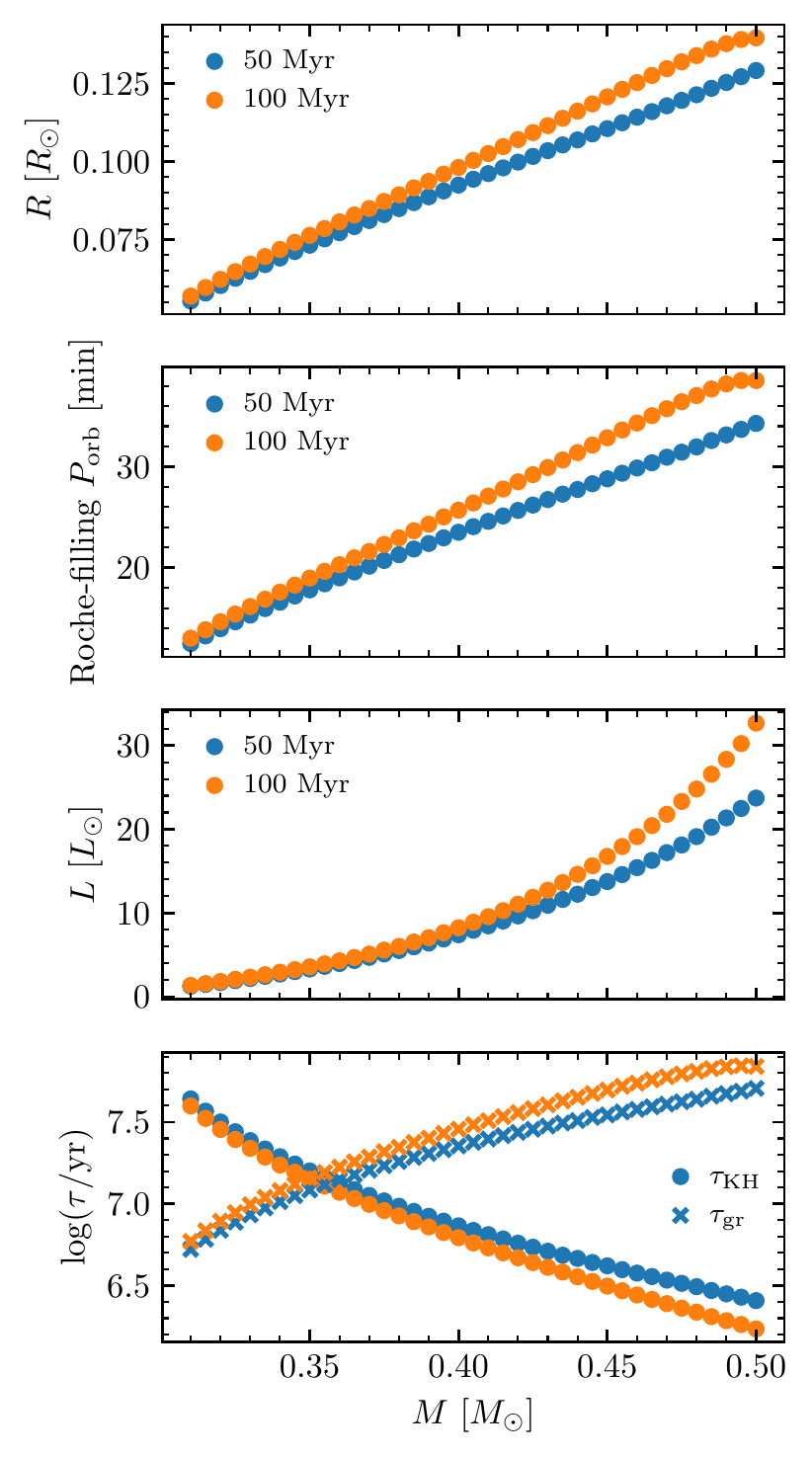}
  \caption{Properties of MESA models for bare Helium stars with a $0.75\ M_\odot$
    companion. The timescale $\tau_{\rm gr}$ in the bottom panel is
    calculated using the Roche-filling $P_{\rm orb}$ from the second
    panel.}
  \label{fig:HeMR}
\end{figure}

By the time the hydrogen envelope is completely removed at an orbital period on
the order of $P_{\rm orb} \approx 30\ \rm min$ for the canonical
0.47~$M_\odot$ sdB model, the timescale for GWR
to shrink the orbit is much shorter at $\tau_{\rm gr} \approx 38\ \rm Myr$.
Figure~\ref{fig:HeMR} shows the mass-radius relation for He cores
based on a grid of MESA models with no H envelope. This
relation gives $\zeta_{\rm eq} \approx 2$ for He stars with no H envelope,
and therefore Equations~\eqref{eq:zetaRLest} and~\eqref{eq:mdot_full}
give $\dot M_1 \approx 10^{-8}\ M_\odot\,\rm yr^{-1}$ when He mass
transfer begins. The second panel of Figure~\ref{fig:HeMR} shows the
periods at which sdB He cores will fill the Roche lobe with a
0.75~$M_\odot$ companion, calculated
with Equations~\eqref{eq:EggletonRL} and~\eqref{eq:Kepler3} by setting
$R_{RL} = R$ using the radius from the top panel.

Figure~\ref{fig:HeMR} also shows the mass-luminosity
relation for these cores, which allows calculating $\tau_{\rm KH}$.
The much lower luminosity and radius of sdB stars with
$M_1 \lesssim 0.35\ M_\odot$ results in the orbital evolution
timescale becoming shorter than the Kelvin-Helmholtz timescale, so sdB
stars that lose mass beyond this point can no longer maintain thermal
equilibrium and will respond to mass loss adiabatically. The index to
describe the radius response for $M_1 \lesssim 0.35\ M_\odot$ is
therefore $\zeta_{\rm ad}$ rather than $\zeta_{\rm eq}$. We can
approximate this index as $\zeta_{\rm ad} = -1/3$ for an ideal gas
polytrope \citep{Soberman1997}, meaning that the adiabatic response of
the star is to expand slightly as it loses mass. Stable mass transfer therefore
requires that $\zeta_{RL} < -1/3$, or $q \lesssim 2/3$ according to
Equation~\eqref{eq:zetaRLest}, but this only requires that the WD
companion have mass $M_2 > 0.525\ M_\odot$, so this condition will
generally be satisfied. Lower mass sdB stars, or those that lose
enough mass to enter this adiabatic regime, also have shorter orbital
periods (2nd panel in Figure~\ref{fig:HeMR}) corresponding to
$\tau_{\rm gr}$ as short as 5--10~Myr (4th panel).
With $\zeta_{\rm ad} - \zeta_{RL} \approx 0.5$,
Equation~\eqref{eq:mdot_full} therefore predicts mass transfer rates
of order $\dot M_1 \gtrsim 10^{-7}\ M_\odot\,\rm yr^{-1}$ in this lower mass
regime.

The general picture for He transfer then is that a canonical
0.47~$M_\odot$ sdB star should begin transferring He at $\dot M_1
\approx 10^{-8}\ M_\odot\,\rm yr^{-1}$, and the mass transfer rate will
gradually increase up to a value approaching $10^{-7}\ M_\odot\,\rm yr^{-1}$
as the sdB star loses mass, which agrees with previous work modeling
mass transfer from He stars \citep{Savonije1986,Iben1991,Yungelson2008,Brooks2015}.

\section{Post Common-Envelope Periods and Contact Scenarios}
\label{s.periods}

\begin{figure}
  \centering
  \includegraphics{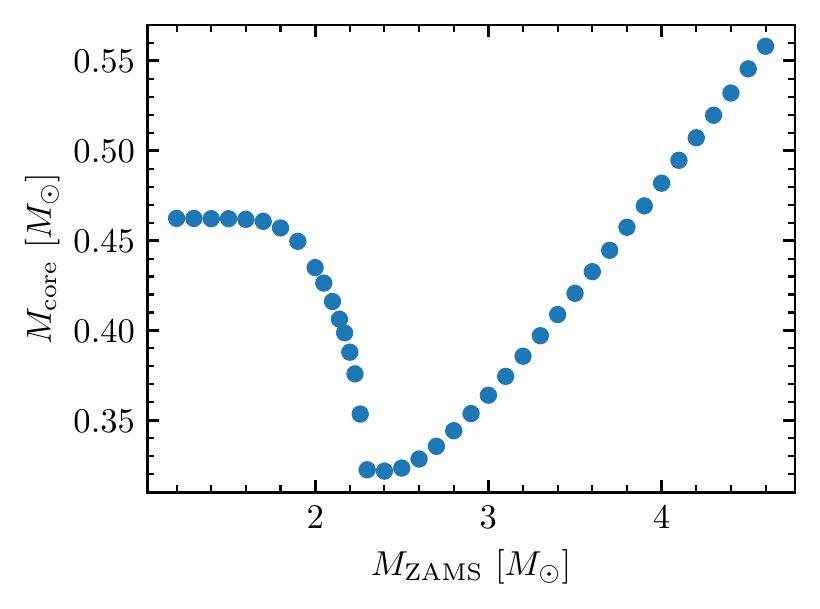}
  \caption{He-burning core masses for a range of ZAMS masses. Stars
    with $M_{\rm ZAMS} \lesssim 2.3\ M_\odot$ undergo the He core flash
    to ignite He off-center, while more massive stars ignite He under
    non-degenerate conditions in the center.}
  \label{fig:CoreMass}
\end{figure}

In this section, we show the range of post-common envelope
binary (PCEB) orbital periods that can lead to mass transfer from a
hot subdwarf to a white dwarf companion.
After a PCEB exits the common envelope at period $P_{\rm init}$,
it spirals inward toward shorter orbital periods. If the
  GWR inspiral timescale at $P_{\rm init}$ is shorter than the
  He-burning lifetime of the sdB star, the binary will eventually make
  contact and transfer mass from the sdB to the WD.
On the other hand, if the orbital period is long enough that the
inspiral time is longer than the He-burning lifetime, the sdB star
will exhaust its nuclear fuel, and the binary will merge later as a
double WD system that includes a ``hybrid'' He/C/O WD \citep{Zenati2019,Schwab2021}.
Using models for sdB radius and lifetime,
we can find the maximum value of $P_{\rm init}$
for which a system can make contact as an sdB+WD.
In order to map out this range of orbital periods, we start with
a grid of MESA models run from ZAMS to construct a variety of
He-burning core masses. Figure~\ref{fig:CoreMass} shows the relation
between zero-age main-sequence (ZAMS) mass $M_{\rm ZAMS}$ and the
resulting He core mass $M_{\rm core}$ for models with no overshoot and
initial metallicity $Z=0.02$.
  The $M_{\rm core}$--$M_{\rm ZAMS}$ relation is sensitive at the
  $\approx$10\% level to parameters like initial metallicity and
  overshoot (through changes to opacities, convection, and mixing),
  but our grid here is sufficient for our purposes in that it produces
  a representative range of He-burning core masses for sdB models.
A more detailed investigation of the $M_{\rm core}$--$M_{\rm ZAMS}$
relation for models including a range of metallicities and overshoot
treatments can be found in \cite{Ostrowski2020}.

\begin{figure*}
  \centering
  \includegraphics{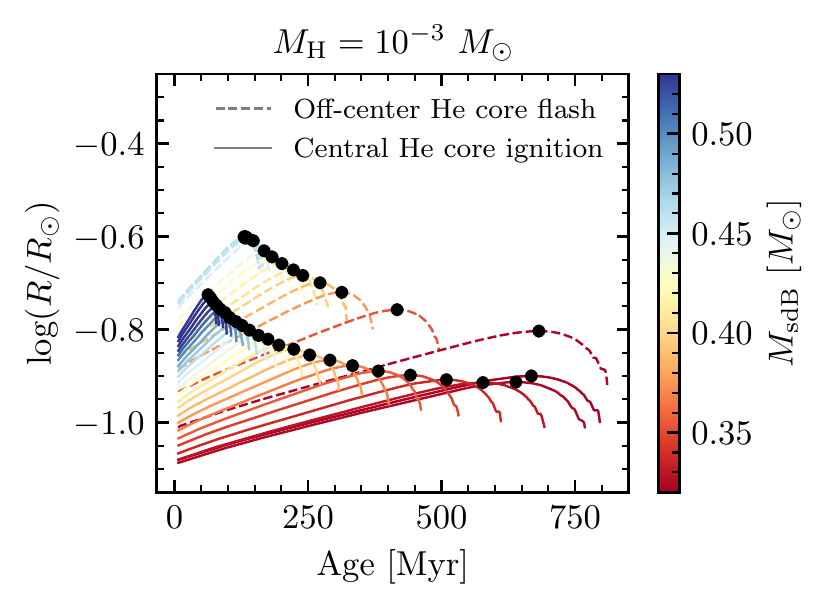}
  \includegraphics{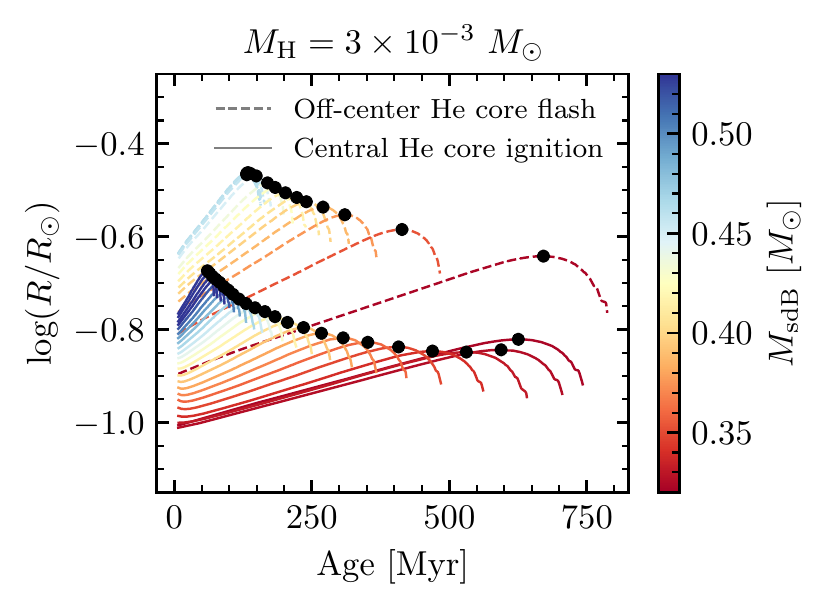}
  \includegraphics{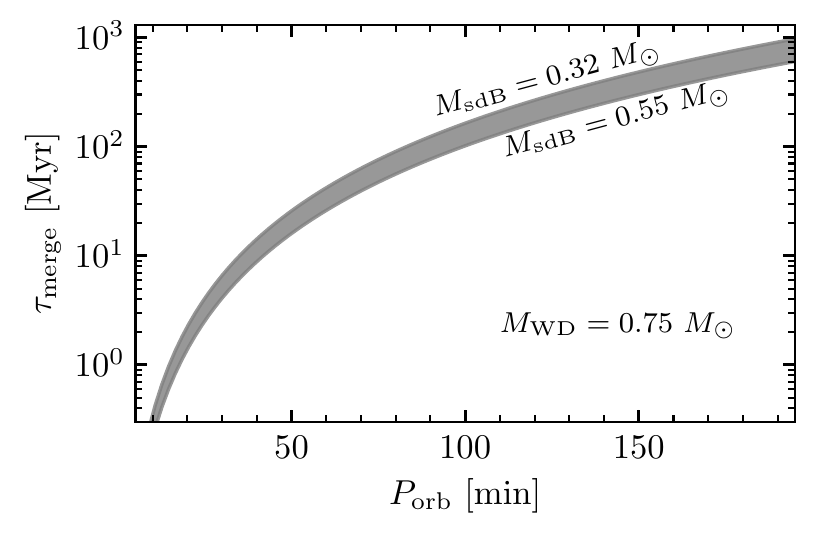}\\
  \includegraphics{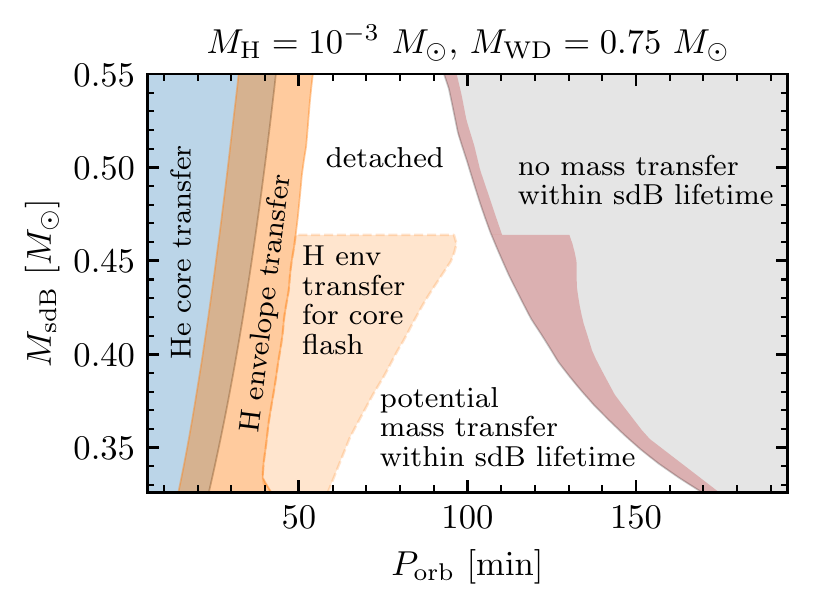}
  \includegraphics{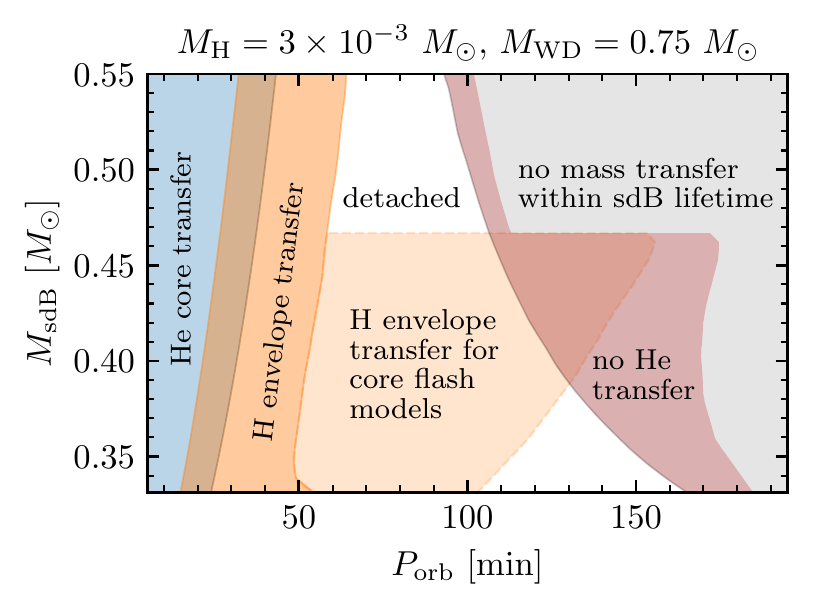}
  \caption{
    {\it Upper Panels:} Radius evolution through the end of core He
    burning for a grid of hot subdwarf
    models with masses in the range $M_{\rm sdB} = 0.32$--$0.55\
    M_\odot$, with $M_{\rm H} = 10^{-3}\ M_\odot$ in the left panel
    and $M_{\rm H} = 3 \times 10^{-3}\ M_\odot$ in the right panel.
    Black points mark the location on each track where the radius
    reaches its maximum value during core He burning.
    Dashed curves indicate models for which
    the progenitor mass was $M_{\rm ZAMS} \leq 2.3\ M_\odot$ and the
    He core ignited off-center. Solid curves indicate models for which
    the progenitor mass was $M_{\rm ZAMS} > 2.3\ M_\odot$ and the He
    ignition occurred under non-degenerate conditions at the center.
    {\it Middle Panel:} Merger time $\tau_{\rm merge}$ calculated
    using Equation~\eqref{eq:tmerge} for sdB stars with masses in the
    range $M_{\rm sdB} = 0.32$--$0.55\ M_\odot$ and a 0.75~$M_\odot$
    WD companion.
    {\it Lower Panels:} Orbital period ranges over which different
    phases of mass transfer can occur. Orange shaded regions show
    orbital period ranges over which the H envelope can transfer to
    the WD companion, calculated using the radii from the upper
    panels. The blue shaded region shows the orbital period range
    where the orbit can become compact enough to transfer material
    from the He core of the sdB star. The gray shaded region to the
    right shows orbital periods for which the inspiral time is too
    long for the system to transfer any mass within the sdB star's He
    burning lifetime. The red shaded region shows the range of orbital
    periods for which the system may be able to inspiral close enough
    to transfer some of the H envelope, but will not have time to
    become compact enough for mass transfer to reach the He core.
  }
  \label{fig:Porb}
\end{figure*}

After the models ignite He, we strip the outer H envelope down to a
specified total H mass (either $10^{-3}\ M_\odot$ or $3 \times
10^{-3}\ M_\odot$) and evolve the model as an sdB star through its core
He-burning lifetime. Note that the total envelope mass can vary
depending on how much He is contained in the H-rich layers above the
He core, and in general $M_{\rm env} > M_{\rm H}$. For the stars
descended from progenitors with $M_{\rm ZAMS} \lesssim 2.3\ M_\odot$
that go through the He core flash, there is a sharp boundary between
the He core and the H envelope, so that the envelope has roughly solar
composition and its total mass is $M_{\rm env} \approx M_{\rm H}/0.7$.
On the other hand, sdB stars descended from progenitors with
$M_{\rm ZAMS} \gtrsim 2.3\ M_\odot$ that ignite He in the center have
envelope material that has been partially burned as part of the
receding convective core earlier on the main sequence, so that the
mass fraction of hydrogen in the region of the star that becomes the
sdB envelope is $X < 0.7$, and it can be as low as $X \sim 0.2$
\citep{Kupfer2020a}. This
envelope composition affects the compactness of the envelope in our
sdB models and hence the overall compactness of the sdB stars. For
example, a $0.47\ M_\odot$ sdB star can descend either from a MS star
of mass $M_{\rm ZAMS} = 1.0$--$1.5\ M_\odot$ or from a higher mass
MS star with $M_{\rm ZAMS} \approx 3.8\ M_\odot$. For our sdB models,
the star with the higher mass progenitor has a much more He-rich
envelope and a more compact overall structure, as shown in radius
evolution displayed in the upper panels of Figure~\ref{fig:Porb}.

Using these sdB models, we then calculate the maximum PCEB orbital
period $P_{\rm init}$ for which an sdB binary system will be able to
spiral inward and transfer mass during its core He burning lifetime.
For each model, we take the maximum radius during core He burning
(identified by the black points in the upper panels of Figure~\ref{fig:Porb}),
and we calculate the orbital period at which it will fill its Roche
lobe using Equations~\eqref{eq:EggletonRL} and~\eqref{eq:Kepler3},
assuming a WD companion mass of 0.75~$M_\odot$. We then
use the age for the maximum radius of each model in the upper panels
of Figure~\ref{fig:Porb} to calculate the maximum initial
orbital period for which the system will
have time to spiral inward and reach contact according to
Equation~\eqref{eq:tmerge}. We use a similar grid of bare He models
with no H envelope to identify the analogous contact and initial
periods for mass transfer from the more compact He cores.

The shaded regions in the lower panels of Figure~\ref{fig:Porb} map
out the orbital period ranges over which H and He phases of mass
transfer can occur, as well as the range of $P_{\rm init}$ for which
an sdB+WD system will not be able to spiral into contact with the sdB
star's He core burning lifetime. For example, a $0.47\ M_\odot$
sdB star with a sufficiently thick H-rich envelope will be able to
fill the Roche lobe and transfer some hydrogen for $P_{\rm init}$ as
long as 175~min, but it can only spiral in far enough to transfer
material from the He core for $P_{\rm init} < 105\ \rm min$.
It will first fill its Roche lobe and begin transferring the H
envelope at an orbital period in the range $P_{\rm orb} =
50$--$150$~min. It will eventually exhaust the envelope and begin
transferring material from the He core at $P_{\rm orb} < 35$~min.
The WD companion mass can also have a minor impact on the
  orbital period ranges displayed in these plots. See the Appendix for
  more discussion on this effect.

\section{MESA Binary Evolution Models}
\label{s.MESAbinaries}

To illustrate typical sdB+WD binary evolution scenarios, we run two
example MESA
binary models through the mass transfer phases up to the point of He
thermonuclear runaway on the accreting WD. These models calculate the
mass transfer rate according the implicit Roche lobe overflow scheme
described in \cite{Paxton2015} for the \cite{Ritter1988} mass transfer
prescription, and they therefore provide independent verification for
the analytic accretion rate estimates from Section~\ref{s.analytics}.

The angular momentum in these binary evolution models is driven by GWR
losses as described by Equations~\eqref{eq:Jdotgr}
and~\eqref{eq:Jorb}. We evolve both the sdB donor star and a
0.75~$M_\odot$ WD accretor, ignoring rotation in both stars for
simplicity. Our modeling for the accreting WD closely follows
\cite{Bauer2017}, including the
$^{14}{\rm N}(e^-, \nu) {^{14}{\rm C}} ( \alpha, \gamma) {^{18}{\rm O}}$
(NCO) reaction chain that may initiate 
the eventual $3\alpha$ thermonuclear runaway depending on the
accretion rate.
The ${^{14}{\rm C}} ( \alpha, \gamma) {^{18}{\rm O}}$ rate in our MESA
models is that of \cite{Johnson2009} as in \cite{Bauer2017}.
\cite{Neunteufel2017,Neunteufel2019}
have argued that rapid rotation and magnetic fields due to the angular
momentum from the accreted material on the WD may significantly alter
the thermal structure of the WD's accreted envelope and modify the
eventual thermonuclear runaway in the accreted He. However, we interpret
the calculations of \cite{Fuller2014} as indicating that dynamical tides and
dissipation due to WD pulsation modes should prevent the WD from
spinning up to rotation rates near critical. Instead, tides are likely to
drive the WD toward much slower rotation rates on the order of
$P_{\rm orb}$, which would be far too slow to effect the envelope
structure of a compact WD with dynamical time $t_{\rm dyn} \ll P_{\rm orb}$.
This is also consistent with the fact that observations rule out rapid
rotation for the accreting WD in some AM CVn systems \citep{Roelofs2006,Kupfer2016}.
Therefore, it should be safe to ignore rotation in
calculating the structure of the He envelope accreted onto the WD.

  As the sdB spirals inward toward its companion, tides may cause it
  to spin up to rotation on the order of the orbital period, though
  tidal dissipation may not be efficient enough to bring the donor star into fully
  synchronous rotation \citep{Preece2018}. Even for fully synchronous
  rotation, we can use Equation~\eqref{eq:Kepler3} to find that a
  Roche-filling donor would rotate with a frequency
  $\Omega_{\rm orb}^2/\Omega_{\rm crit}^2 = (1+1/q)(R_{RL}/a)^3$,
  where $\Omega_{\rm crit} = \sqrt{GM_{1}/R_{1}^3}$ is the
  critical rotation rate of the donor star and $\Omega_{\rm orb} =
  2\pi/P_{\rm orb}$. Using Equation~\eqref{eq:EggletonRL} along with
  the above formula, we find that
  $\Omega_{\rm orb}^2/\Omega_{\rm crit}^2 \approx 0.1$
  for any $q \lesssim 1$, so distortion
  in the outer layers due to rotation would have only a minor
  impact on the donor sdB star. Since this is an upper limit on the
  rotation rate due to potential tidal spin-up, we feel it is
  justified to neglect the structural effects of rotation for the MESA
  sdB models in this section.

We assume conservative mass transfer ($\dot M_2 = - \dot M_1$)
unless the WD is undergoing a nova outburst that causes
it to expand and overflow its Roche lobe. In that case, we adopt the
following simple prescription for removing mass from the WD to allow
it to evolve through nova cycles while staying within its Roche lobe
and allowing binary evolution to continue. When the WD
radius expands beyond 0.85~$R_{RL,2}$ (where $R_{RL,2}$ is the WD
Roche radius), we use a simple exponential Roche lobe overflow scheme
for the WD to remove mass from system at a rate of
\begin{equation}
  \dot M_2 = (10^{-5}\ M_\odot\,{\rm yr}^{-1})
  \exp \left( \frac{R_2 - 0.85 R_{RL,2} }{3 \times 10^{-3}\ R_\odot}\right)~.
  \label{eq:WDmdot}
\end{equation}
This mass is assumed to be lost from the WD as a spherical wind, and
therefore carries away specific angular momentum matching the WD orbit.
It should be noted that the parameters in Equation~\eqref{eq:WDmdot}
are chosen for simplicity rather than based on any physical
motivation.
These parameters were chosen by experimenting to find values
  that allowed the WD to expand near its Roche lobe and then quickly
  ramp up mass loss to allow robustly evolving through multiple
  complete nova cycles, but they are not necessarily unique in
  accomplishing this goal.
The details of this WD mass loss scheme therefore should
not be construed as physical, but the small amount of mass lost from
the system during these nova cycles will not have a large impact on
the overall orbital evolution.
For a more thorough discussion of the impact of mass loss
prescriptions on MESA models of novae, see \cite{Wolf2013}.
For simplicity, we also
do not consider any interaction between the donor and mass lost
from the WD during these hydrogen novae, which may lead to some
dynamical friction that enhances the rate at which the system loses
angular momentum and spirals inward, though this process is uncertain
(e.g., \citealt{Shen2015}).

\begin{figure*}
  \centering
  \includegraphics{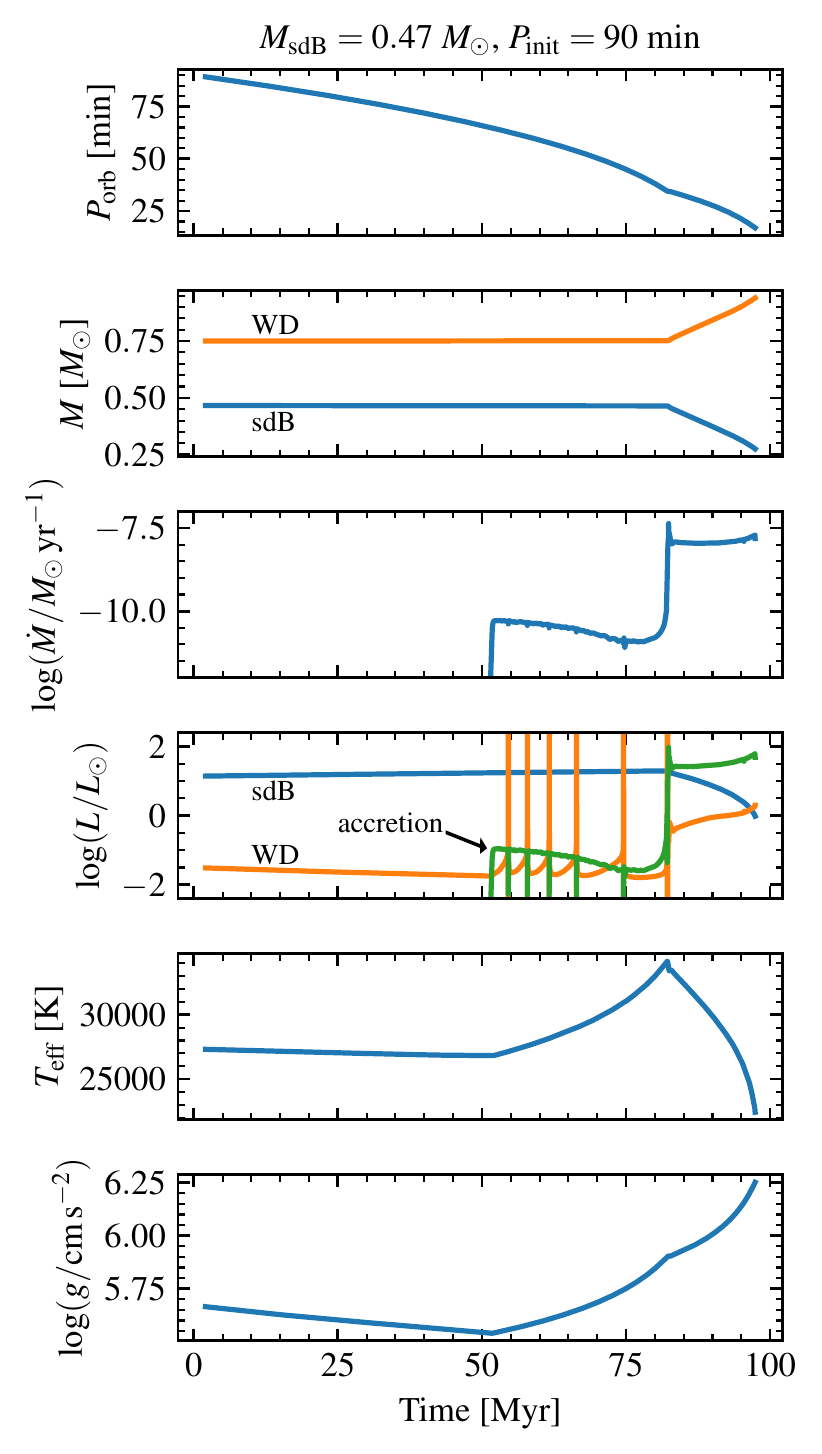}
  \includegraphics{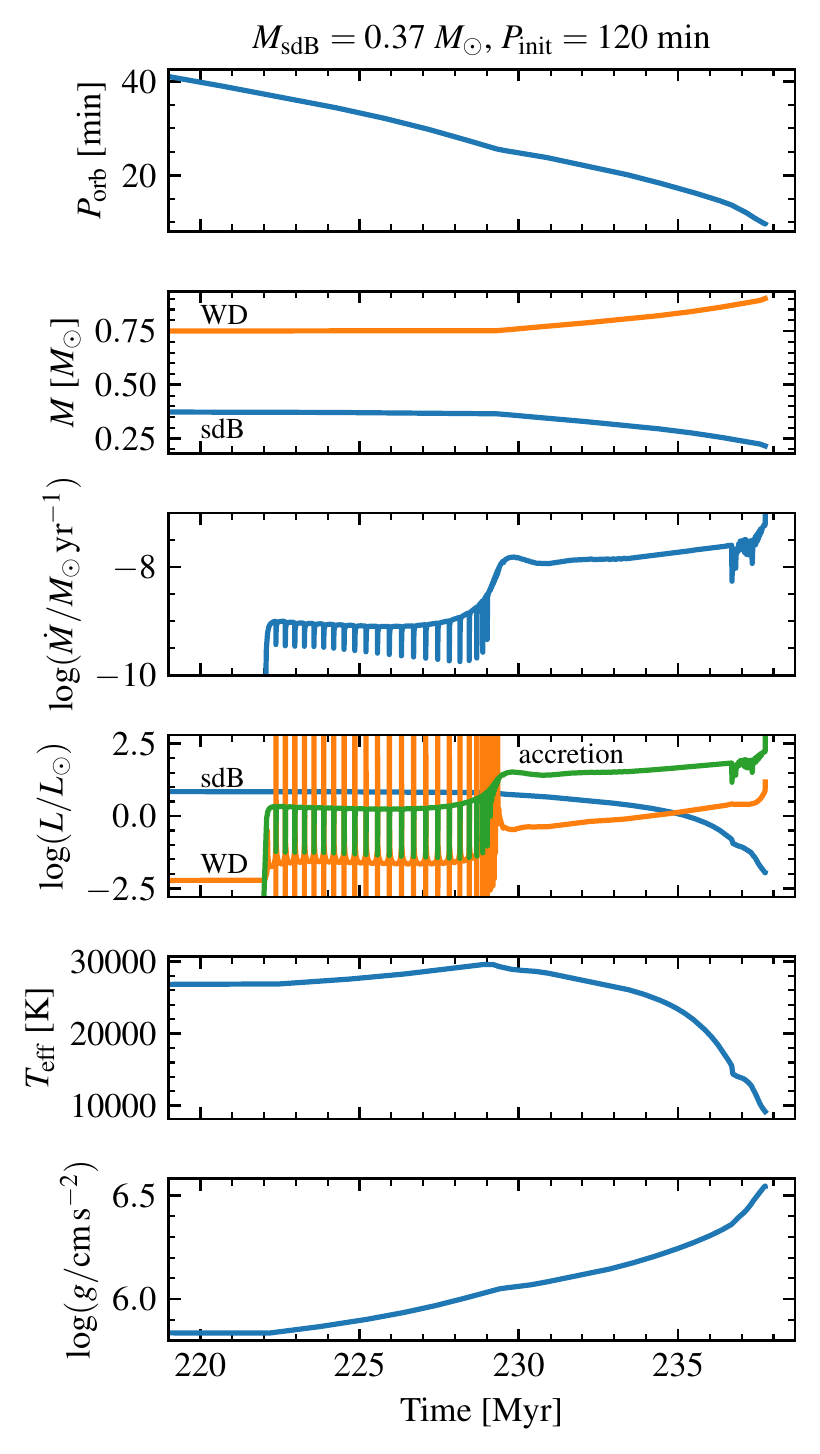}
  \caption{
    MESA binary evolution models of sdB stars with a 0.75~$M_\odot$ WD
    companion. The left panels show the model for a canonical
    0.47~$M_\odot$ sdB star described in Section~\ref{s.canon}, and
    the right panels show the model for a 0.37~$M_\odot$ model
    described in Section~\ref{s.0.37}. The x-axis in these plots is
    time since He core burning started in the sdB model, which also
    corresponds to when the envelope was removed from the progenitor
    star to form a subdwarf, presumably through a common envelope that
    also leaves the binary systems at the initial orbital periods
    $P_{\rm init}$ labeled at the top of the plots.
  }
  \label{fig:MESAbinary}
\end{figure*}

\subsection{Canonical Mass sdB Donor}
\label{s.canon}

Our canonical sdB model is 0.47~$M_\odot$ with an envelope mass of
$M_{\rm env} = 10^{-3}\ M_\odot$, descended from a 1.0~$M_\odot$ model
that has the H envelope removed during the He core flash. We
initialize this in a binary with orbital period $P_{\rm init} = 90\
\rm min$ with a 0.75~$M_\odot$ WD companion. The left panel of
Figure~\ref{fig:MESAbinary} shows that this system comes into contact
after 50~Myr when $P_{\rm orb} = 65\ \rm min$,
and donates the H envelope at a rate of $\dot M \lesssim
10^{-10}\ M_\odot\, {\rm yr}^{-1}$ for 30~Myr. During this time, the
WD accretor model undergoes 6 H novae while accreting H-rich material,
and ejects most of the $10^{-3}\ M_\odot$ of accreted material from
the system. The hot subdwarf donor increases its $T_{\rm eff}$ to nearly
$35{,}000$~K as the Roche lobe shrinks and the hot subdwarf becomes more
compact while maintaining a nearly constant luminosity before finally
exposing the He core.

Once the He core is exposed and begins to overflow the Roche lobe
when $P_{\rm orb} = 34\ \rm min$, the
mass transfer rate increases to $\dot M \sim 10^{-8}\ M_\odot\, {\rm
  yr}^{-1}$ for another 20 Myr, in agreement with the prediction of
Section~\ref{s.analytics}. The sdB donates nearly 0.2~$M_\odot$ of
He-rich material to the WD companion, and its luminosity and $T_{\rm
  eff}$ decrease as core He burning subsides and eventually shuts off
altogether in the donor star, while the remaining He core continues to
grow more compact. Eventually, the WD accumulates enough He that the
NCO chain initiates a dynamical thermonuclear runaway at the base of
the accreted He envelope, likely leading to a detonation.
The density at the ignition location in the envelope of this
  model is $\rho = 1.7 \times 10^6\,\rm g\,cm^{-3}$, above the
  critical density threshold $\rho_{\rm crit} = 10^6\,\rm g\,cm^{-3}$ for which a
  detonation is likely \citep{Woosley1994,Woosley2011,Bauer2017,Neunteufel2017}.
This detonation may destroy the entire WD and liberate the
remnant of the sdB donor to become a runaway star with a velocity
close to its final orbital velocity of $780\ \rm km\, {\rm s}^{-1}$
\citep{Neunteufel2019,Bauer2019,Neunteufel2020}.

The fourth panel in Figure~\ref{fig:MESAbinary} gives bolometric
luminosities of both the sdB donor and WD accretor as well as the
estimated accretion luminosity $L_{\rm acc} \approx GM_{\rm WD}\dot
M/R_{\rm WD}$. This shows that the sdB star dominates the luminosity
of the system before mass transfer and during the envelope transfer
phase (except during novae), but the accretion luminosity may become
comparable or larger during the He transfer phase. Late in the
evolution of these systems, the accretion disk may dominate observed
luminosity in both the ultraviolet and optical.
In this phase, these binaries
would appear as AM Canum Venaticorum (AM CVn) systems, with negligible
luminosity from the sdB donor once it loses enough mass to drop below
$\approx$0.3~$M_\odot$. This particular model donates some carbon
along with the He during this phase due to most of the He core
containing some ashes from the He core flash, and the spectral
signatures of this C in the accretion disk would be atypical for an AM
CVn. Donors descended from higher mass progenitors that ignite He in
their centers rather than through an off-center core flash do not
contain any C ashes outside the convective He-burning core. These sdB
stars contain a $\approx$0.2--0.3~$M_\odot$ shell of unprocessed
He-dominated material in their outer layers to transfer during the He
overflow phase, and this would form a more typical AM CVn spectrum
from the accretion disk, showing nitrogen features from the ashes of
main sequence CNO burning, but not carbon features.
See also \cite{Yungelson2008} for discussion of the composition of
material donated from He stars corresponding to the scenario of
higher mass progenitors that did not experience an off-center He
flash.

\subsection{0.37 $M_\odot$ sdB Donor}
\label{s.0.37}

Our second binary evolution model contains a 0.37~$M_\odot$ sdB star
descended from a $M_{\rm ZAMS} = 3.0\ M_\odot$ progenitor (cf.\ 
Figure~\ref{fig:CoreMass}). The envelope contains a total hydrogen
mass of $M_{\rm H} = 10^{-3}\ M_\odot$ like the models in the left
panel of Figure~\ref{fig:Porb}. However, the H mass fraction in the
layers that form the eventual sdB envelope is only $X \approx 0.2$ due to
prior nuclear processing in the receding convective core during the MS
evolution. This leads to a very compact overall structure for the sdB
model because of the higher He content in the envelope, and the total
mass of material in the outer layers that contain some H is
$M_{\rm env} = 7 \times 10^{-3}\ M_\odot$. We initialize this sdB
model in a binary with a 0.75~$M_\odot$ WD with an orbital period of
$P_{\rm init} = 120\ \rm min$. It first comes into contact after
220~Myr when the orbital period has decreased to $P_{\rm orb} = 37\ 
\rm min$, and it then begins donating the envelope material at a rate
of $\dot M \sim 10^{-9}\ M_\odot\,{\rm yr}^{-1}$. Over the next 7~Myr,
the WD accretor undergoes 22 H novae and ejects much of the accreted
envelope mass from the system. The downward spikes in the accretion
rate in the right panel of Figure~\ref{fig:MESAbinary} are due to
temporary widening of the orbit due to this small amount of mass loss
during each nova cycle. These downward spikes are a direct result of
our oversimplified prescription for removing mass from the WD during
novae as a simple spherical wind, so the orbital evolution briefly
following each nova is likely oversimplified and unphysical as well.
However, this has little impact on the overall
orbital evolution and mass transfer, as each nova removes only
$\sim 10^{-4}\ M_\odot$ from the system, and GWR quickly brings the system
back into contact with a similar equilibrium mass transfer rate to
that before the nova occurred.

Eventually  when $P_{\rm orb} = 26\ \rm min$
the underlying He core is exposed, and mass then begins to transfer to
the WD at a rate on the order of $10^{-8}\ M_\odot\,{\rm yr}^{-1}$ for
the next 9~Myr. The system transfers a total of 0.15~$M_\odot$ before
$3\alpha$ burning in the degenerate accreted He leads to a
thermonuclear runaway in the WD envelope. The final donor velocity in
this model is 950~$\rm km\,s^{-1}$, but the ignition location of the WD
envelope occurs in less dense layers
($\rho = 4 \times 10^5\,\rm g\,cm^{-3}$) outward from the base of the
accreted He due to a temperature inversion \citep{Brooks2015,Bauer2017}, so
this particular model is unlikely to lead to a detonation that
liberates the donor as a runaway star. If the WD accretor does not
detonate, this system could continue transferring mass and evolve
through a period minimum
around $P_{\rm orb} = 10\ \rm min$ when the donor is
  $\approx$0.2~$M_\odot$ \citep{Yungelson2008,Neunteufel2020},
eventually transferring hybrid
He/C/O material from the partially burned core as the system evolves
back toward longer orbital periods \citep{Nelemans2010}.

\section{Detectability of Ellipsoidal Variation}
\label{s.ellipsoidal}

Ellipsoidal modulation of the lightcurves in sdB+WD binary systems is
the key observable that will enable discovery of new systems based on
time-domain observations with enough epochs or short enough cadence to resolve orbital
periods on the order of 20~minutes to 3~hours. In order to demonstrate
the detectability of this modulation, in this section we use
{\tt LCURVE}\footnote{\url{https://github.com/trmrsh/cpp-lcurve}}
\citep{Copperwheat2010} to calculate the amplitude of
variation by modeling the lightcurve variability of ellipsoidally
deformed sdB stars in binary systems with WD companions.
We model phase folded lightcurves with 1000 data points.
We use values for limb darkening and gravity darkening
based on \cite{Claret2017} and assume a typical sdB star with
$T_{\rm eff} = 25{,}000\,\rm K$ and $\log(g/{\rm cm\,s^{-2}}) = 5.5$.
We treat the WD companion as a point mass, so that there are no
eclipses as we are only interested in the ellipsoidal amplitude.
The geometry of the ellipsoidally deformed sdB star is the primary
source of flux variation. Limb darkening and gravity darkening have
small influence on the ellipsoidal amplitude in our models (order 1\%
level) but can increase towards hotter temperatures.
Doppler beaming can also slightly enhance the overall
  amplitude by boosting one of the lightcurve peaks relative to the
  other peak (e.g., \citealt{Bloemen2011}), as seen in \cite{Pelisoli2021} with
  TESS observations of HD~265435. For actively accreting systems, an
  accretion disk can further alter the lightcurve through the
  Rossiter--McLaughlin effect or due to irradiation, as seen in
  \cite{Kupfer2020a}.
Our models in
this section therefore represent a conservative estimate of the total
amplitude of variation in an ellipsoidal system, as the amplitude can
be further enhanced by eclipse features from a WD or disk at high inclination
or greater limb and gravity darkening for a subdwarf with
$T_{\rm eff} > 25{,}000\,\rm K$.

\begin{figure}
  \centering
  \includegraphics{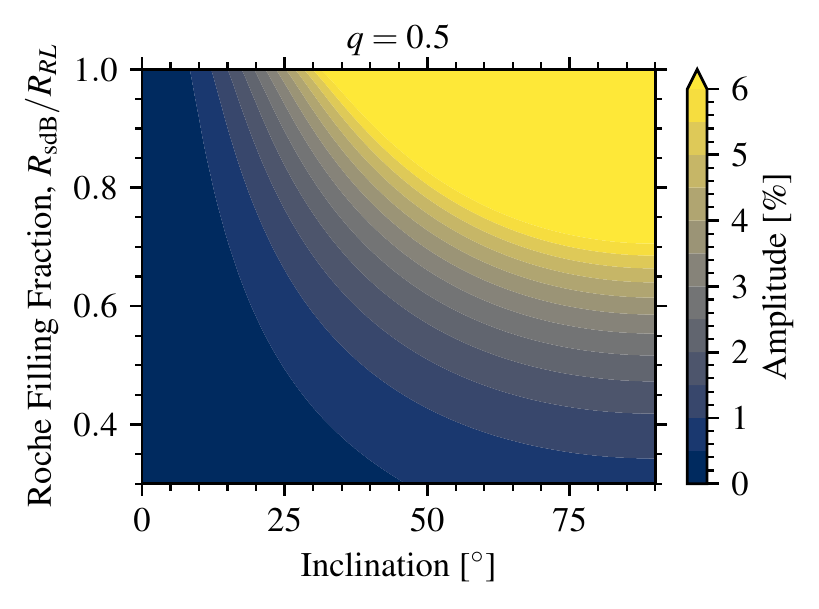}
  \includegraphics{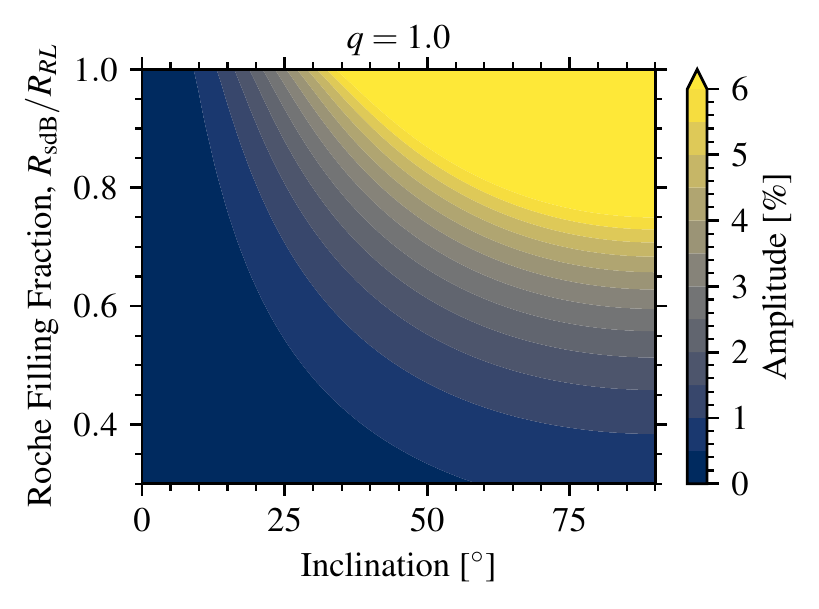}
  \caption{
    Contours showing the observable amplitude of flux variability as a
    function of orbital inclination and Roche-filling fraction for two
    different mass ratios: $q = 0.5$ (upper panel) and $q = 1.0$
    (lower panel). Variability amplitudes are calculated using {\tt
      LCURVE} \citep{Copperwheat2010} assuming a typical sdB star with
    $T_{\rm eff} = 25{,}000\,\rm K$ and $\log(g/{\rm cm\,s^{-2}}) = 5.5$.
  }
  \label{fig:Amplitude}
\end{figure}

We find that the main factors influencing the amplitude of ellipsoidal modulation
are the Roche-filling fraction of the sdB ($R_{\rm sdB}/R_{RL}$)
and inclination (see equation~4 in \citealt{Bloemen2012}).
Figure~\ref{fig:Amplitude} therefore shows the amplitude of
ellipsoidal modulation as a function of these two variables. The mass
ratio can also have a minor impact on the amplitude, so we show the
amplitude for two different mass ratios ($q=0.5$ and $q=1.0$).

As an example to illustrate the detectability of ellipsoidal variation in practice,
for sdB stars within 2~kpc the variability amplitudes in
Figure~\ref{fig:Amplitude} translate into the following
sensitivity to detection by the Zwicky Transient Facility (ZTF,
\citealt{ZTF1,ZTF2}), which was used to discover the systems of
  \cite{Kupfer2020a,Kupfer2020b}.
Assuming an sdB star with absolute magnitude of 4 (typical in the {\it Gaia}
G band) yields an apparent magnitude of 15.5 at 2~kpc without any
extinction. Stars in the Galactic plane may experience an additional
0.7--1.0~mag of extinction per kpc \citep{Krelowski1993,Green2019}, yielding 17--17.5~mag for stars at 2~kpc.
For stars at this magnitude, ZTF has a flux precision of
$\Delta F/ F \sim1.5-2\%$ (see figure~11 in \citealt{Masci2019}).
Therefore, a photometric amplitude of 5--6\%
can be easily detected with a $>3\sigma$
significance. The contours in Figure~\ref{fig:Amplitude} therefore
show that all Roche-filling sdB stars with inclination
$i \gtrsim 30^\circ$ should be easily detectable at this distance, and
a significant fraction of sdB stars should be detectable even when
they have only reached 70--80\% of Roche-lobe filling.

In the future, the Legacy Survey of Space and Time (LSST, \citealt{LSST}) of the Vera
Rubin Observatory will enable pushing this detection sensitivity out
to even further distances (several kpc) and smaller
Roche-filling fractions for sdB+WD systems. A magnitude of 17
is the bright end for LSST, and the photometric
precision goal is 5~mmag in g and r (0.5\%; see Table~14 in the
requirements document%
\footnote{\url{https://www.lsst.org/scientists/publications/science-requirements-document}}).
An amplitude of just 1.5--2\% will therefore be
easily detected with $>3\sigma$ significance, enabling detection of
some sdB systems that are only $\approx$50\% of Roche-filling
according to Figure~\ref{fig:Amplitude}. For a typical compact
sdB+WD system that is born by exiting a common envelope at $P_{\rm
  orb}\approx 1.5-3\ \rm hours$, the sdB could spend up to 100s of Myr
with detectable ellipsoidal deformation as it spirals inward before
filling its Roche lobe and beginning to transfer mass for another few
10s of Myr. Detection of the distribution of sdB+WD binaries with
ellipsoidal modulation in LSST could therefore map out nearly the
entire lifetime of this phase, from birth to eventual mass transfer,
supernova, or merger.

\section{Discussion and Conclusions}
\label{s.conclusions}

We have calculated characteristic timescales and orbital periods for
mass transfer from sdB stars to WD companions. Our results imply that
many undiscovered compact sdB+WD binary systems with Roche-filling or
near Roche-filling sdB donors have yet to be discovered. We can estimate a
lower limit for the number of observable systems based on the following
argument. \cite{Kupfer2020a,Kupfer2020b} found two short-period (39
and 56~min) subdwarf+WD binaries within 2 kpc in which the subdwarf is currently
overflowing its Roche lobe and transferring mass to the WD. These
systems were discovered because the subdwarf donors are very hot
($T_{\rm eff} > 33{,}000\ \rm K$), which allowed discovery due to
ellipsoidal modulation in a sample with strict color cuts even after
reddening due to dust in the Galactic plane where these systems
reside \citep{Kupfer2020b}.

Modeling in \cite{Kupfer2020a,Kupfer2020b} indicated that these observed
systems are currently evolving through a short-lived $\approx$1~Myr
phase in which the subdwarf becomes hotter just after core He burning has
ended. Similar systems that exited a common envelope at slightly
shorter orbital periods would come into contact Myr earlier while the
donor is still a cooler He burning sdB star. Our calculations in this
paper indicate that the mass transfer phases for both the outer H
envelope and the He core last for $\sim$10~Myr when the binary comes
into contact while the sdB star is still burning He in its
core.
Section~\ref{s.ellipsoidal} shows that the variability of these
systems will be easily detectable by ZTF for all Roche-filling systems
within 2~kpc except those with small inclinations $i \lesssim 30^\circ$.
Therefore, at least 10s of binary systems with Roche-filling
donors in both of these phases should be observable, and there
are likely 100s more that are close enough to Roche-filling
to exhibit detectable ellipsoidal modulation according to Figure~\ref{fig:Amplitude}. Clearly, many more
sdB+WD systems await discovery among the millions of variable stars
observed in the Galactic plane \citep{Kupfer2021}.

Most of these systems have not yet been discovered because they likely
reside in the Galactic plane, where the strict color cuts of previous
searches for sdB stars have resulted in excluding these systems
due to reddening. The demographics of the few systems that we do know
of so far
\citep{Vennes2012,Geier2013,Kupfer2017a,Kupfer2017b,Kupfer2020a,Kupfer2020b}
indicate that they come from a young population in
the Galactic plane. Indeed, measured WD companion masses are often
significantly larger than 0.6~$M_\odot$, indicating short MS
progenitor lifetimes prior to the common envelope. The population
synthesis models of \cite{Han2003} also indicate that sdB+WD binaries
with $P_{\rm orb} < 3\ \rm hr$ primarily descend from stars with
$M_{\rm ZAMS} > 2\ M_\odot$ that do not experience a degenerate core
He flash, another indicator that these systems must descend from a
young population.

Additionally,
  \cite{Geier2019} present a catalog of $\approx$40,000 sdB candidates
  from Gaia DR2, but their selection criteria are likely to exclude
  many sdB stars in the Galactic plane associated with a young
  population in dense stellar regions.
  Their selection criteria include quality metrics from
  Gaia DR2 which remove blended sources and sources with bad
  astrometry. This affects in particular dense stellar regions, which
  makes the selection based on the \cite{Geier2019} catalog very
  incomplete at low Galactic latitudes with high stellar
  densities. Therefore, searches based on this catalog will miss a
  significant number of sdBs from a young population.
sdB stars descended from such a young population 
would have the more compact envelope configurations of stars that do
not experience the He core flash in Figure~\ref{fig:Porb}, and
therefore they should typically experience the onset of H envelope
mass transfer at orbital periods of 40--60~minutes according to the
darker orange shaded region in the lower panels of
Figure~\ref{fig:Porb}. We therefore predict that this period range
should be the most fruitful in searching to discover new Roche-filling
sdB+WD binary systems.

Discovering and characterizing this population of sdB+WD binary
systems will provide useful constraints on common envelope
outcomes. The lower panels of Figure~\ref{fig:Porb} show that systems
which spiral close enough to transfer mass from the sdB star must exit
the common envelope at orbital periods shorter than the shaded
regions on the right side of those plots. Modeling of well-characterized
systems will therefore provide a direct link to post-common-envelope
outcomes for this class of binary systems.

Extending our estimates to the total number of systems in the Galaxy
based on the volume in which observed systems have been detected so far,
we estimate that our Galaxy contains at least $\sim 10^3$--$10^4$
sdB+WD binaries with orbital periods shorter than 1--2 hours. This lower limit is based on the two systems of
\cite{Kupfer2020a,Kupfer2020b} that were discovered within a distance
of $d \approx 2\, \rm kpc$ using ZTF in the Northern Hemisphere. We can then scale the 10s to 100s of
observable systems that we estimate should exist within a few kpc by a factor of
$M_\star/[(\pi/2) d^2 \Sigma_d] \sim 100$ to arrive at the total number of systems
in the Galaxy, where $\Sigma_d$ is the local stellar surface density and 
$M_\star$ is the total Galactic stellar mass \citep{McGaugh2016}.
These sdB+WD systems will therefore be a significant source of
Galactic foreground gravitational waves for LISA, and some may be
resolvable LISA sources. Our estimates of the stars in this population
are closely related to the work of \cite{Gotberg2020}, who focused on
He stars in binaries with
slightly longer orbital periods $P_{\rm orb} \geq 60\ \rm min$ just
outside the range of where their He star models may begin to fill
their Roche lobes and transfer mass. Our models therefore represent a
subset of the population of $\approx 10^5$ such systems that they calculate,
with the evolution extended toward shorter orbital periods to continue
through mass transfer.

Many of the sdB+WD systems with $P_{\rm orb} \lesssim 2\ \rm hr$ are
likely to produce an eventual thermonuclear explosion of a thick
($\sim$0.1~$M_\odot$) He shell on the accreting WD. With typical
inspiral times on the order of $10^8$~yr, our estimate of
$10^3$--$10^4$ systems therefore predicts that the rate of such
explosions in our Galaxy could be as high as one per $10^4$~yr,
comparable to the rate of ``.Ia'' supernovae predicted from double WD
AM CVn systems \citep{Bildsten2007,Shen2009}.

We have also found that during the He transfer phase, the accretion
luminosity is often the dominant source of luminosity in the system,
and therefore many of these systems may spend several Myr in a state
that appears very similar to the properties of AM CVn systems
\citep{Nelemans2001,Nelemans2004,Marsh2004}.
For sdB donors descended from $M_{\rm ZAMS} \gtrsim 2.3\ M_\odot$
progenitors that do not experience the He core flash, the outer
$\approx$0.2~$M_\odot$ of He-dominated material that transfers onto
the WD has not undergone any He-burning to process the CNO abundances,
and so the composition of accretion disks in these systems would be
indistinguishable from the He WD donor channel for AM CVns
\citep{Nelemans2010b}.
Therefore we argue that it is possible that a significant number of
relatively short-period (10~min$\,< P_{\rm orb} <\,$30~min) AM
CVn-like systems associated with a young stellar population in the
Galactic plane may be descended from sdB+WD binaries, even though the
donor composition may point towards CNO cycle material that has
traditionally been interpreted as indicating a He WD donor.

In fact, one such AM CVn-like system may already be known.
SDSS~J190817.07+394036.4 is an AM CVn system \citep{Fontaine2011} with
an orbital period of 18~minutes, and \cite{Kupfer2015} found from the relation between the superhump and the orbital period that
this system likely has a mass ratio of $q=0.33$, which is
problematic for a double WD interpretation of this system.
According to \cite{Marsh2004}, a double WD system with this mass ratio
would likely experience unstable mass transfer and merge, while we
have found that $q<2/3$ sdB+WD systems will experience stable mass
transfer. Our binary model in Section~\ref{s.0.37} evolves through
stable He mass transfer with a mass ratio and orbital period very
similar to those observed for SDSS~J1908.

In principle, these systems would also be distinguishable from the
classic He WD AM CVn scenario by the direction of their orbital period
evolution. The sdB+WD systems we discuss in this paper have
$\dot P_{\rm orb} < 0$, evolving toward shorter orbital periods, while
AM CVn systems with He WD donors spend most of their lifetime evolving
in the opposite direction toward longer orbital periods. A detection
of negative $\dot P_{\rm orb}$ could therefore serve as another
indicator that an AM CVn system comes from the sdB+WD binary
channel. We expect orbital period changes on the order of
$\dot P_{\rm orb} = -(3/8)P_{\rm orb}/\tau_{\rm merge}$ when
negligible mass is being transferred so that we can use
Equations~\eqref{eq:Kepler3} and~\eqref{eq:tmerge} directly to
calculate a purely GWR-driven value for $\dot P_{\rm orb}$.
This equation gives orbital period
evolution rates of $|\dot P_{\rm orb}| \lesssim 10^{-12}$--$10^{-11}\ \rm s\,s^{-1}$
for sdB+WD binary systems. He mass
transfer from the less massive sdB to the more massive WD slows
orbital evolution to make $|\dot P_{\rm orb}|$ smaller
relative to purely GWR-driven binary evolution (cf.\
Equation~\eqref{eq:Jorb} and the top panels in
Figure~\ref{fig:MESAbinary}). In practice, the magnitude of this
orbital period evolution may be detectable after a few years of
regular monitoring, at least for some eclipsing systems with precise
timing.

\begin{acknowledgments}

We thank the anonymous referee for an insightful report that led to many
valuable improvements to this manuscript, including much of the
material in Section~\ref{s.ellipsoidal} in particular.
We thank Tom Marsh for valuable discussions that originated ideas
about sdB hydrogen envelope transfer that grew into this work.
We thank Josiah Schwab and the other organizers of {\it The Beginnings
  and Ends of Double White Dwarfs} hosted at the Niels Bohr Institute
in Copenhagen during July 2019, where those discussions were
stimulated. This work also benefited from the subsequent workshop held
at DARK in July 2019 that was funded by the Danish National Research
Foundation (DNRF132).
We thank Warren Brown, Lars Bildsten, Tom Marsh, and Patrick Neunteufel for comments and
suggestions that improved this manuscript as it was in preparation.
This research benefited from interactions that were funded by the
Gordon and Betty Moore Foundation through grant GBMF5076.
This work was supported by the National Science Foundation through
grants PHY-1748958 and ACI-1663688.

\end{acknowledgments}

\software{
\texttt{MESA} \citep[][\url{http://mesa.sourceforge.net}]{Paxton2011,Paxton2013,Paxton2015,Paxton2018,Paxton2019},
\texttt{MESASDK} 20.12.1 \citep{mesasdk_linux,mesasdk_macos},
\texttt{LCURVE} \citep[][\url{https://github.com/trmrsh/cpp-lcurve}]{Copperwheat2010}
\texttt{matplotlib} \citep{hunter_2007_aa}, 
\texttt{NumPy} \citep{der_walt_2011_aa,Harris2020}
}

\appendix

\section{Impact of White Dwarf Companion Mass}
\begin{figure*}
  \centering
  \includegraphics{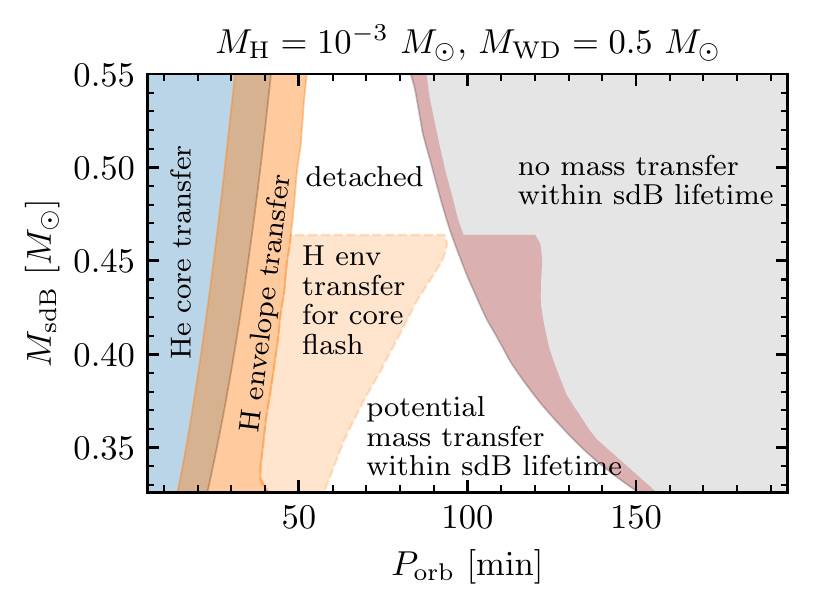}
  \includegraphics{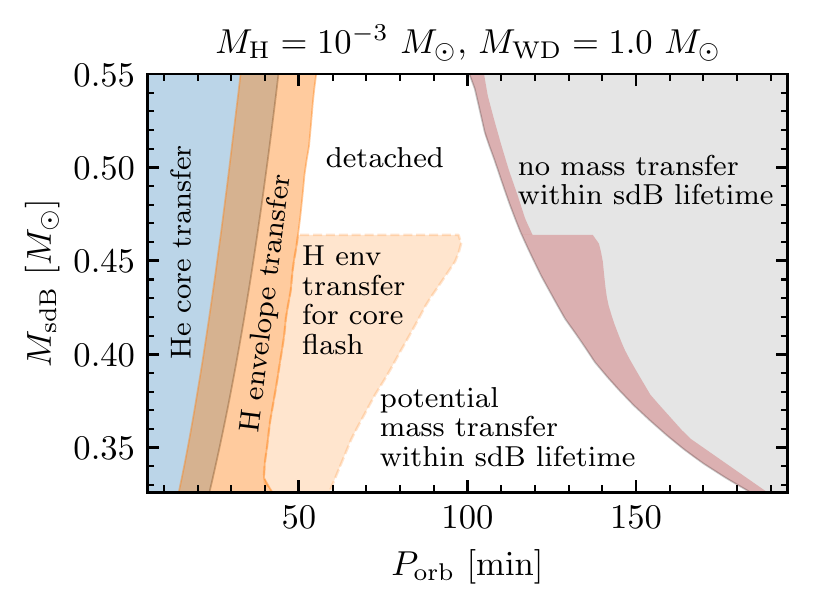}
  \caption{
    Same as the lower left panel of Figure~\ref{fig:Porb}, except with
    different WD accretor masses (0.5~$M_\odot$ in the left panel and
    1.0~$M_\odot$ in the right panel) to show how companion mass can
    cause minor shifts to the period ranges where these systems may be
    detected.
  }
  \label{fig:PorbAppendix}
\end{figure*}

Figure~\ref{fig:PorbAppendix} shows the small impact of the WD companion
mass on the orbital period ranges over which the different phases of
mass transfer can occur as calculated in Section~\ref{s.periods}. The
WD companion mass enters this analysis in two ways. First, it changes
the size of Roche radius for a given sdB star mass, but as
Figure~\ref{fig:PorbAppendix} shows this effect is very small when
translated into orbital period space. Second, the WD companion mass
changes the gravitational inspiral timescale according to
Equation~\eqref{eq:tmerge}, which has a noticeable impact on the
orbital period range at which the PCEB system can exit the common
envelope and still be able to make contact within the sdB star
He-burning lifetime. More massive WD companions cause this inspiral
timescale to be faster, moving the boundary for $P_{\rm init}$
required for a system to make contact out toward longer orbital
periods.

%\clearpage

\bibliography{refs}
\bibliographystyle{yahapj}

\end{document}